\documentclass[a4paper]{iopart}
\usepackage{subeqn} 
\usepackage[square,sort&compress]{natbib}

\newcommand{\bk}{{\bf k}} 
 
\newcommand{\bt}{{\bf t}}
\newcommand{\bx}{{\bf x}} 
\newcommand{\by}{{\bf y}}
\newcommand{\bz}{{\bf z}}

\newcommand{\bS}{{\bf S}}
\newcommand{\bT}{{\bf T}}

\newcommand{\kbt}{{k_{\rm B}T}} 

\begin{document}
\title{Anisotropic glass-like properties in tetragonal disordered crystals}
\author{D V Anghel$^{1}$ and D. Churochkin$^{2,3}$}
\address{$^1$ Department of Theoretical Physics, National Institute for Physics and Nuclear Engineering--''Horia Hulubei'', Bucharest - Magurele, Romania\\
$^2$ R\&D Institute 'Volga', Saratov 410052, Russia\\
$^3$ Saratov State University, 410012, Astrakhanskaya St. 83, Saratov, Russia}

\begin{abstract}
The low temperature acoustic and thermal properties of amorphous,
glassy materials are remarkably similar. All these properties are 
described theoretically with reasonable quantitative 
accuracy by assuming that the amorphous solid contains dynamical defects 
that can be described at low temperatures as an ensemble of two-level systems 
(TLS), but the deep nature of these TLSs is not clarified yet. 
Moreover, glassy properties were found also in disordered crystals, 
quasicrystals, 
and even perfect crystals with a large number of atoms in the unit cell. 
In crystals, the glassy properties 
are not universal, like in amorphous materials, and also exhibit anisotropy. 
Recently it was proposed a model for the interaction of two-level systems with 
arbitrary strain fields (Phys. Rev. B {\bf 75}, 64202, 2007), which was 
used to calculate the thermal properties of nanoscopic membranes at 
low temperatures. The model is also suitable for the description of 
anisotropic crystals. We describe here the results of the calculation of 
anisotropic glass-like properties in crystals of various lattice symmetries, 
emphasizing the tetragonal symmetry. 
\end{abstract}

\section{Introduction} \label{intro}

The low temperature acoustic and thermal properties of amorphous,
glassy materials are remarkably
similar and they can be explained--to a large extent--by assuming that
the material contains a large number of dynamic defects. These dynamic
defects are tunneling systems (TS) which are modeled by an ensemble of
two-level systems (TLS) 
\cite{JLowTempPhys.7.351.1972.Philips,PhilMag.25.1.1972.Anderson}.
Crystals with defects--with a large enough amount of
disorder--exhibit also glass-like properties, but these properties
are not so universal and, even more, they are not
isotropic--for example the absorption and the propagation velocity change 
of elastic waves depend on the propagation and polarization directions 
of the wave \cite{PhysRevB.51.8158.1995.Laermans}. 

A microscopic model of the tunneling systems is still missing. 
For this reason, the study of disordered crystals
is especially interesting because it offers additional
information: in some materials we know
quite well which are the entities that tunnel between different
equilibrium positions. Beside this, the anisotropy of the
TLS-sound wave interaction in crystals represents another
challenge to the interaction models of TLSs which requires
clarification.

A free TLS is described in a two-dimensional Hilbert space by the Hamiltonian 
\begin{equation}
H_{\rm TLS} = \frac{\Delta}{2}\sigma_z + \frac{\Lambda}{2}\sigma_x 
\equiv \frac{1}{2}\left(\begin{array}{cc}\Delta&\Lambda\\-\Lambda&\Delta
\end{array}\right)
\end{equation}
where $\sigma_z$ and $\sigma_x$ are Pauli matrices; the 
parameters $\Delta$ and $\Lambda$ are called the \textit{asymmetry of 
the potential} and \textit{tunnel splitting}, respectively. The excitation 
rate of the TLS, $\epsilon$, is obtained by diagonalizing $H_{\rm TLS}$ 
and this gives $\epsilon=\sqrt{\Delta^2+\Lambda^2}$. 

The perturbation of the free TLS, induced by a strain field is 
$H_{\rm I}\equiv(\delta/2)\sigma_z$, where in general 
$\delta\equiv 2\gamma_{ij} S_{ij}$, with $S_{ij}$ being the strain tensor and 
$\gamma_{ij}$ a symmetric tensor that characterizes the TLS and 
its interaction with the strain field--here, as everywhere in this paper, 
we assume \textit{summation over repeated indices}. In bulk isotropic 
materials, like in amorphous solids, the elastic waves may be decomposed into 
longitudinally and transversely plane waves and the 
interaction term $\delta$ may be written as 
\begin{equation}
\delta= 2\gamma_\sigma S_\sigma , \label{delta_iso}
\end{equation}
where $\sigma$ is the polarization, 
$l$ (longitudinal) or $t$ (transversal), of the perturbing wave. 
In this case the absorption rate of a phonon of polarization $\sigma$ and 
wavevector $\bk$, by a TLS is 
\begin{equation}
\Gamma_{\bk,\sigma} = \frac{2\pi}{\hbar}\gamma_\sigma^2 \frac{\hbar k}{2V\rho c_\sigma}\frac{\Lambda^2}{\epsilon^2}n_{\bk,\sigma} \delta(\hbar c_\sigma k-\epsilon) , 
\label{Gammaabs}
\end{equation}
where by $n_{\bk,\sigma}$, $\rho$, $c_\sigma$, and $V$ we denote the population 
of the phonon mode, the density of the solid, the sound velocity and the volume 
of the solid, respectively. 
If the number of TLSs of asymmetry and tunnel splitting in the intervals 
$\rmd\Delta$ and $\rmd\Lambda$ is $P(\Delta,\Lambda)\rmd\Delta\rmd\Lambda=VP_0/\Lambda\rmd\Delta\rmd\Lambda$, 
then integrating $\Gamma_{\bk,\sigma}$ over the TLS ensemble and using 
$n_{\bk,\sigma}=[\exp(\beta\hbar c_\sigma k)-1]^{-1}$, we obtain the 
\textit{standard} total phonon absorption rate, 
\begin{equation}
\left(\tau^{\rm (S)}_{\bk,\sigma}\right)^{-1}
= P_0\gamma^2_\sigma K\cdot\tanh\left(\frac{\epsilon}{2\kbt}\right), 
\label{av_STM}
\end{equation}
where we introduced the short-hand notation 
$K\equiv \pi n_{\bk\sigma}k/(V\rho c_\sigma)$. 

Although widely used, the simplified version of the tunneling model 
exposed above cannot account for the anisotropy of the glass-like 
properties observed in disordered crystals. To explain this anisotropy, 
in Refs. \cite{PhysRevLett.2008.Anghel,EPL.2008.Anghel} we applied 
the model introduced in \cite{PhysRevB75.064202.2007.Anghel}. 
In this model, a direction $\hat\bt$ is associated to 
each TLS [$\hat\bt\equiv(t_x,t_y,t_z)^t$, where the superscript $^t$ 
denotes the transpose of a matrix or vector] and $\delta$ was 
written in the form \cite{PhysRevB75.064202.2007.Anghel,PhysRevB76.165425.2007.Kuhn,JPhysConfSer.92.12133.2007.Anghel}
\begin{equation}
\delta=2\bT^t\cdot[R]\cdot\bS, 
\end{equation}
with 
$\bT\equiv(t_x^2,t_y^2,t_x^2,2t_yt_z,2t_zt_x,2t_xt_y)^t$ and $\bS$ 
is the strain tensor in abbreviated subscript notations. 
The $6\times6$ symmetric matrix $[R]$ contains the TLS-strain field 
coupling constants $r_{ij}$ and its structure is determined by 
the symmetries of the lattice in which the TLS is embedded. 
In Refs. \cite{PhysRevLett.2008.Anghel,EPL.2008.Anghel} we 
applied this model to calculate the glassy properties of disordered 
cubic (Ca stabilized zirconium \cite{Topp:thesis}), trigonal (neutron 
irradiated quartz \cite{PhysRevB.51.8602.1995.Keppens}) and hexagonal crystals. 
Here we shall report results on tetragonal lattices. 

\section{Phonon scattering rates in tetragonal lattices}

In the new model, the phonon absorption rate by a TLS is 
\begin{equation}
\Gamma_{\bk\sigma}(\hat\bt) =
\frac{2\pi}{\hbar}\frac{\Lambda^2n_{\bk\sigma}}
{\epsilon^2}|\bT^t\cdot[R]\cdot\bS_{\bk\sigma}|^2\delta(\epsilon-\hbar\omega).
\label{eqn_Gamma_bar}
\end{equation}
We see that the main characteristic of the TLS-phonon interaction is 
contained in the quantity
$M_{\bk,\sigma}(\hat\bt)\equiv\bT^t\cdot[R]\cdot\bS_{\bk\sigma}$, which bears 
an intrinsic anisotropy through the matrix $[R]$, on which the
symmetries of the lattice are imposed \cite{PhysRevB75.064202.2007.Anghel,JPhysConfSer.92.12133.2007.Anghel,PhysRevLett.2008.Anghel,EPL.2008.Anghel}
To calculate the average scattering rate of a phonon by the
ensemble of TLSs, now we have to average not only over the distribution 
of $\Delta$ and $\Lambda$, but also over the distribution of
$\hat\bt$. In this way we get the total phonon absorption rate, 
\begin{equation} 
\tau^{-1}_{\bk\sigma} =
\frac{2\pi P_0\tanh\left(\frac{\epsilon}{2\kbt}\right)}{\hbar} 
n_{\bk\sigma}\langle|M_{\bk\sigma}(\hat\bt)|^2\rangle.
\label{av_def}
\end{equation}
To reduce the number of degrees of freedom of the
problem, in what follows we shall assume that $\hat\bt$ is
isotropically oriented.

For a tetragonal lattice of symmetry classes \textit{4mm, 422,
$\overline{4}$2m, 4/mmm}, the matrix $[R]$ has the form
\cite{Auld:book}
\begin{eqnarray}
[R] = \left(\begin{array}{cccccc}
r_{11}&r_{12}&r_{13}&0&0&0\\
r_{12}&r_{11}&r_{13}&0&0&0\\
r_{13}&r_{13}&r_{33}&0&0&0\\
0&0&0&r_{44}&0&0\\
0&0&0&0&r_{44}&0\\
0&0&0&0&0&r_{66}
\end{array}\right), && \label{R_trigonal}
\end{eqnarray}
similar to that of the tensor of elastic stiffness constants,
$[c]$.
The difference between the trigonal lattice of symmetry
\textit{32} and the chosen classes of tetragonal lattice is that
$r_{14}$ and $c_{14}$ are zero. In that sense the hexagonal and
tetragonal lattices are similar to each other. However, in
contrast to both hexagonal and trigonal \textit{32}
systems, for the tetragonal lattice the condition
$r_{66}=(r_{11}-r_{12})/2$ does not hold anymore. As a
consequence, the number of independent constants is equal to that
for trigonal \textit{32} lattice, whereas the structure of
$[R]$ is similar to that for the hexagonal lattice.  

For concreteness, we shall consider the class \textit{422}, with a 
system of coordinates chosen such that 
the $z$ and $x$ axes are the 4-fold and
2-fold rotational symmetry axes, respectively, while the $y$ axis
is perpendicular to both $x$ and $z$. Solving the Christoffel
equation we find that the crystal can sustain pure
transversal waves in all three directions, $x$, $y$, and  $z$.
For the transversal waves propagating in the $x$
direction, the ones polarized in the $z$ direction propagate with
the velocity $\sqrt{c_{44}/\rho}$ whereas the ones polarized in
the $y$ direction propagate with the velocity
$\sqrt{c_{66}/\rho}$. The transversal waves propagating in the $y$
direction are similar to the ones propagating in the $x$
direction: the waves polarized in the $x$ direction have a sound
velocity of $\sqrt{c_{66}/\rho}$, whereas the ones polarized in
the $z$ direction have a sound velocity of $\sqrt{c_{44}/\rho}$.
Finally, the transversal waves propagating in the $z$ direction
have all the same sound velocity, $\sqrt{c_{44}/\rho}$.

Averaging over the directions $\hat\bt$ we obtain for the transversely 
polarized waves propagating in the $\hat\bx$,$\hat\by$ and $\hat\bz$ directions
the results 
\begin{subequations}\label{av_defT}
\begin{eqnarray}
\langle|M_{k\hat\bx,\hat\by,t}|^2\rangle=\langle|M_{k\hat\by,\hat\bx,t}|^2\rangle=\frac{4N^2k^2r_{66}^2}{15}
\label{av2_def}
\end{eqnarray}
and 
\begin{equation}
\fl
\langle|M_{k\hat\bx,\hat\bz,t}|^2\rangle=\langle|M_{k\hat\by,\hat\bz,t}|^2\rangle=
\langle|M_{k\hat\bz,\hat\bx,t}|^2\rangle=\langle|M_{k\hat\bz,\hat\by,t}|^2\rangle=\frac{4N^2k^2r_{44}^2}{15}.
\label{av3_def}
\end{equation}
\end{subequations}
in obvious notations: the first subscript indicates the
propagation direction and the second denotes the direction of polarization. 
Plugging equations (\ref{av2_def}) and (\ref{av3_def}) into (\ref{av_def}) 
we obtain the general type of expression 
\begin{equation}
\left(\tau^{\rm (T)}_{\bk,\sigma}\right)^{-1}
= P_0(\gamma^{\rm (T)}_{\bk,\sigma})^2 K\cdot\tanh\left(\frac{\epsilon}{2\kbt}\right), 
\label{av_STM1}
\end{equation}
where the superscript $^{\rm (T)}$ comes from the tetragonal symmetry. 
Concretely, the new $\gamma$ constants are 
\begin{subequations} \label{gammasH}
\begin{eqnarray}
(\gamma^{\rm (T)}_{k\hat\by,\hat\bx,t})^2 &=&
(\gamma^{\rm (T)}_{k\hat\bx,\hat\by,t})^2
= \frac{4r_{66}^2}{15}, \label{gammaYXtH} \\
(\gamma^{\rm (T)}_{k\hat\bx,\hat\bz,t})^2 &=&
(\gamma^{\rm (T)}_{k\hat\by,\hat\bz,t})^2 =
(\gamma^{\rm (T)}_{k\hat\bz,\hat\bx,t})^2 =
(\gamma^{\rm (T)}_{k\hat\bz,\hat\by,t})^2 = \frac{4r_{44}^2}{15} .
\label{gammaXZtH}
\end{eqnarray}
\end{subequations}
The ratio $r_{66}/r_{44}$ can be obtained by measuring 
the characteristics (sound velocity change or attenuation rate) of 
elastic waves propagating along the $\langle100\rangle$ crystallographic 
directions. It is interesting to note now the similarity between relations 
like $\tau^{\rm (T)}_{k\hat\bx,\hat\bz,t}/\tau^{\rm (T)}_{k\hat\bx,\hat\by,t}=(\gamma^{\rm (T)}_{k\hat\bx,\hat\by,t}/\gamma^{\rm (T)}_{k\hat\bx,\hat\bz,t})^2=(r_{66}/r_{44})^2$ and relations like $(c_{k\hat\bx,\hat\by,t}/c_{k\hat\bx,\hat\bz,t})^2=c_{66}/c_{44}$ that hold for tetragonal lattices--e.g. $c_{k\hat\bx,\hat\by,t}$ is the velocity 
of a transversal sound wave propagating along the $x$ direction and 
polarized along the $y$ direction.


%

\section{Conclusions}

We calculated the average scattering rates of pure 
transversal phonon modes on the TLSs in tetragonal disordered crystals, 
to emphasize the anisotropy of the glass-like properties imposed by
the lattice anisotropy. 
Two of the parameters of the model--the TLS-strain field coupling 
constants--may be obtained by measuring
attenuation rates or sound velocity changes of elastic waves 
propagating along the different $\langle100\rangle$ directions. 
Determining the coupling constants enables one to calculate the 
glassy properties of the crystal in any direction. 

The calculations can be extended easily to disordered crystals of
any symmetry. Moreover, although we used in our calculations an
isotropic distribution over the TLS orientations, the comparison
of our calculations with experimental data would enable one to
find if our assumption is true or not. If it is not true, one can
determine, at least in principle, the distribution of the
orientations of the TLSs.

\ack

This work was partly supported by the NATO grant EAP.RIG 982080. 

\providecommand{\newblock}{}

\end{document}